\documentclass[aps,prl,twocolumn,epsfig,showpacs,preprintnumbers,superscriptaddress,float]{revtex4}
\usepackage{graphicx}
\usepackage{dcolumn}
\usepackage{bm}
\usepackage{amssymb}

\usepackage{amsmath}

\begin{document}
\title
{Anomalous Nernst and Hall effects in magnetized platinum and palladium}
 
\author{G. Y. Guo}
\email{gyguo@phys.ntu.edu.tw}
\affiliation{Department of Physics, National Taiwan University, Taipei 10617, Taiwan}
\affiliation{Graduate Institute of Applied Physics, National Chengchi University, Taipei 11605, Taiwan}
\author{Q. Niu}
\affiliation{International Center for Quantum Materials and Collaborative Innovation Center of Quantum Matter, 
Peking University, Beijing 100871, China} 
\affiliation{Department of Physics, The University of Texas at Austin, Austin, Texas 78712, USA}
\author{N. Nagaosa}
\affiliation{Department of Applied Physics, University of Tokyo, Tokyo 113-8656, Japan}
\affiliation{RIKEN Center for Emergent Matter Science (CEMS), Wako, Saitama 351-0198, Japan}

\date{\today}   

\begin{abstract}
We study the anomalous Nernst effect (ANE) and anomalous Hall effect (AHE) in proximity-induced 
ferromagnetic palladium and platinum which is widely used in spintronics, within the Berry 
phase formalism based on the relativistic band structure calculations. 
We find that both the anomalous Hall ($\sigma_{xy}^A$) and Nernst ($\alpha_{xy}^A$)
conductivities can be related to the spin Hall conductivity ($\sigma_{xy}^S$) and 
band exchange-splitting ($\Delta_{ex}$) by relations $\sigma_{xy}^A =\Delta_{ex}\frac{e}{\hbar}\sigma_{xy}^S(E_F)'$
and $\alpha_{xy}^A = -\frac{\pi^2}{3}\frac{k_B^2T\Delta_{ex}}{\hbar}\sigma_{xy}^s(\mu)''$, respectively. 
In particular, these relations would predict that the $\sigma_{xy}^A$ in the magnetized Pt (Pd)
would be positive (negative) since the $\sigma_{xy}^S(E_F)'$ is positive (negative).
Furthermore, both $\sigma_{xy}^A$ and $\alpha_{xy}^A$ are approximately proportional to the induced
spin magnetic moment ($m_s$) because the $\Delta_{ex}$ is a linear function of $m_s$.
Using the reported $m_s$ in the magnetized Pt and Pd, we predict that the intrinsic anomalous 
Nernst conductivity (ANC) in the magnetic platinum and palladium
would be gigantic, being up to ten times larger than, e.g., iron, while the intrinsic anomalous Hall 
conductivity (AHC) would also be significant. 
\end{abstract}
 
\maketitle

\section{Introduction}
Spin transport electronics (spintronics) has recently attracted enormous attention mainly because of 
its promising applications in information storage and processing and other electronic technologies\cite{Prin98,Zuti04}.
Spin current generation, detection and manipulation are three key issues in the emerging spintronics. 
Large intrinsic spin Hall effect (SHE) in platinum has recently been predicted\cite{Guo08} and 
observed (see Refs. \onlinecite{Feng12} and \onlinecite{Hoff13} and references therein). 
In the SHE, a transverse spin current 
is generated in response to an electric field in a metal with relativistic electron interaction. 
The SHE enables us to generate and control spin current without magnetic field
or magnetic materials, which would be an important step for spintronics. Furthermore, in the inverse spin Hall effect, 
a transverse voltage drop arises due to the spin current\cite{Vale06,Sait06}, and this allows us to 
detect spin current by measuring the Hall voltage. 
Therefore, platinum has been widely used as a spin current generator and detector
in recent spin current experiments, such as spin Seebeck effect\cite{Uchi08},
spin pumping\cite{Kaji10} and spin Hall switching\cite{Miro11}, and plays a unique role
in recent developments in spintronics.

Platinum is an enhanced paramagnet because its 5$d$-band is partially filled 
with a large density of states (DOS) at the Fermi level ($E_F$) [$N(E_F) = \sim$1.74 states/eV/spin].
Consequently, it could become ferromagnetic with a significant spin magnetic moment
when placed next to a ferromagnetic metal\cite{Wilh00,Ante99} or in low-dimensional structures
such as an atomic bilayer on silver (001) surface\cite{Blug95} or a 
freestanding atomic chain\cite{Smog08,Tung10}. Indeed, platinum was reported to possess a magnetic moment 
as large as $\sim$0.2 and $\sim$0.5 $\mu_B$/atom in Ni/Pt 
and Fe/Pt multilayers\cite{Wilh00,Ante99}, respectively.
In a ferromagnetic metal, a transverse charge current would be generated in response
to an electric field due to relativistic spin-orbit coupling (SOC), an effect 
known as the anomalous Hall effect (AHE)\cite{Naga10}, discovered by Hall\cite{Hall81} long ago.
Since the AHE is another archetypal spin-related transport phenomenon\cite{Naga10} 
and the SOC strength in Pt is large, it would be interesting
to study the AHE in the proximity-induced ferromagnetic platinum.
Furthermore, as pointed out in Ref. \onlinecite{Huan12}, the fact that the Hall voltage could
be generated by both the AHE and inverse SHE in the magnetized platinum, might
complicate the detection of the pure spin current and also related phenomena using 
platinum. Therefore, it is important to understand the transport and magnetic properties of 
the magentized platinum.

In a ferromagnet, the Hall voltage could also arise when a thermal gradient instead
of an electric field, is applied. This phenomenon, again due to the relativistic SOC, 
is refered to as the anomalous Nernst effect (ANE)\cite{Nern87}. 
Interestingly, the ANE could be used as a probe of the vortex phase in type II superconductors\cite{Xu00}
and has been receiving considerable attention in 
recent years.\cite{Lee04,Xiao06,Miya07,Berg10,Yoko11,Huan11,Weis13,Schm13}
In this context, it would be interesting to study the ANE in the proximity-induced
ferromagnetic platinum. On the other hand, spin Seebeck effect (another thermal phenomenon), 
which refers to the generation of a spin-motive force in a ferromagnet by a temperature gradient,
has recently attracted considerable attention\cite{Uchi08,Uchi10,Jawo10}. Again, this effect is
usually measured as a transverse voltage in a nonmagnetic metal such as Pt in contact
with the ferromagnet via the inverse SHE\cite{Uchi08}. Clearly, if the metal is magnetized due 
to the magnetic proximity effect, the ANE would contribute to the measured Hall voltage too.
In this connection, it is imperative to understand the ANE in the magnetized platinum.

Palladium is isoelectronic to platinum and thus has an electronic structure
similar to that of Pt except a smaller SOC strength (see, e.g., Refs. \onlinecite{Guo08} and \onlinecite{Guo09}
and references therein). For example, like Pt, Pd also has a large intrinsic
spin Hall conductivity (SHC)\cite{Guo09} and is a highly enhanced paramagnetic metal 
with a large $N(E_F) = \sim$2.69 states/eV/spin. 
In fact, palladium possesses the largest paramagnetic susceptibility of $567\times10^{-6}$
emu/mole among the nonmagnetic metals\cite{Hugu71} and is usually considered to be
nearly ferromagnetic. It could become ferromagnetic
when placed next to a ferromagnetic metal\cite{Celi90,Chil94} or fabricated 
as an atomic bilayer on silver (001) surface\cite{Blug95} or a
freestanding atomic chain\cite{Tung10}. 
Recently, the AHE was observed in the Pd film on an yttrium iron garnet (YIG).\cite{Zhou13}
Surprisingly, it was reported that the intrinsic anomalous Hall conductivity 
(AHC) in the Pd film on the YIG layer has a sign opposite to that for the Pt/YIG bilayer.\cite{Zhou13}
This indicates that the AHC in a magnetized nonmagnetic metal does not simply scale
with the SOC strength. One would then ask what determines the AHC in
the magnetized metals.

In this paper, therefore, we study the AHE and ANE in the proximity-induced
ferromagnetic platinum and palladium within the Berry phase formalism\cite{Xiao10}
based on first-principle relativistic band structure calculations.
We also perform analytic calculations to identify possible relations
between the SHC in an nonmagnetic metal and the AHC in the corresponding magnetized 
metal. The rest of this paper is organized as follows.
In the next section, we briefly describe the Berry phase formalism for 
calculating the AHC and ANC as well as the computational details.
In Sec. III, the calculated AHC and ANC will be presented.
Finally, the conclusions drawn from this work will be summarized in
Sec. IV.

\section{Theory and Computational details}
The anomalous Hall conductivity and anomalous Nernst conductivity (ANC) are calculated
by using the Berry-phase formalism\cite{Xiao10}. Within this Berry-phase formalism, the AHC is 
simply given as a Brillouin zone (BZ) integration of the Berry curvature for all the occupied bands,
\begin{eqnarray}
\sigma_{xy}^{A} = -\frac{e^2}{\hbar}\sum_n \int_{BZ}\frac{d{\bf k}}{(2\pi)^3}f_{{\bf k}n}\Omega_n^z({\bf k}),\nonumber \\
\Omega_n^z({\bf k}) = -\sum_{n'\neq n}
\frac{2{\rm Im}[\langle{\bf k}n|v_x|{\bf k}n'\rangle\langle{\bf k}n'|v_y|{\bf k}n\rangle]}
 {(\epsilon_{{\bf k}n}-\epsilon_{{\bf k}n'})^2},
\end{eqnarray}
where $f_{{\bf k}n}$ and ${\Omega^z_n}$ are the Fermi distribution function and the Berry curvature 
for the $n$th band at ${\bf k}$, respectively.\cite{Yao04} 
Similarly, the ANC can be written as
\begin{eqnarray}
\alpha_{xy}^A = \frac{1}{T}\frac{e}{\hbar}\sum_n \int_{BZ}\frac{d{\bf k}}{(2\pi)^3}f_{{\bf k}n}\Omega_n^z({\bf k})\nonumber \\
\times [(\epsilon_{{\bf k}n}-\mu)f_{{\bf k}n}+k_BT\textrm{ln}(1+e^{-\beta(\epsilon_{{\bf k}n}-\mu)})],
\end{eqnarray} 
where $\mu$ is the chemical potential and $k_B$ is the Boltzmann constant.\cite{Xiao06}

The proximity-induced ferromagnetic platinum and palladium are investigated by the constrained spin-density 
functional theory with the local density approximation to the exchange-correlation potential.\cite{vos80}
Spin-polarized self-consistent scalar-relativistic electronic structure calculations with 
the spin magnetic moment fixed to specified values, are performed. Using the resultant self-consistent 
charge densities, the fully relativistic band structures are then calculated for the AHC and ANC calculations.
The highly accurate all-electron full-potential linearized augmented plane wave (FLAPW) method, as implemented
in the WIEN2K code\cite{wien2k02}, is used. 
The experimental lattice constants $a =3.92$ and 3.89 (\AA) are used, respectively, for Pt and Pd.
In both cases, the muffin-tin sphere radius ($R_{mt}$) of 2.5 a.u. is adopted. 
The wave function, charge density, and potential are expanded 
in terms of the spherical harmonics inside the muffin-tin spheres and the cutoff angular moment ($L_{max}$) 
used is 10, 6 and 6, respectively. The wave function outside the muffin-tin sphere
is expanded in terms of the augmented plane waves (APWs) and a large number of APWs (about 70 APWs per atom, 
i. e., the maximum size of the crystal momentum $K_{max}=9/R_{mt}$) are included in the present calculations. 
The tetrahedron method is used for the BZ integration\cite{Jeps71}. To obtain accurate ground state properties,
a fine 21$\times$21$\times$21 grid of 11616 $k$-points in the first BZ is used.
For the AHC and ANC calculations, a very find grid of 258156 $k$-points on the magnetic irreducible 
wedge (1/16 BZ) in the BZ is used. This is equivalent to a large number of
$k$-points of $\sim 4000000$ in the full BZ, and corresponds to the division of the $\Gamma X$ line into 70 intervals. 
Comparison with test calculations with a denser grid of 381915 $k$-points (80 divisions of the $\Gamma X$ line) 
indicates that the calculated AHC and ANC converge to within a few \%.

\section{Results and discussion}
The relativistic band structure and also AHC ($\sigma_{xy}^A$) as a function of the 
Fermi energy ($E_F$) for the magnetized platinum and palladium with the spin magnetic 
moment $m_s= 0.1$ $\mu_B$/atom are displayed in Figs. 1 and 2, respectively.
All Kramer-degenerate bands in nonmagnetic platinum (see Fig. 1 in \cite{Guo08})
and palladium (see Fig. 1 in \cite{Guo09}) are now exchange-split due to 
the induced magnetization in the magnetized Pt and Pd.
This is clearly visible for the $d$-dominated bands [i.e., energy bands below 1.0 eV in Fig. 1(a) 
or 0.5 eV in Fig. 2(a)] since the ferromagnetism is mainly caused by the exchange interaction among
the $d$-electrons. The band spin-splittings are largest in the flat bands of almost
pure $d$ character such as the bands around 0.5 eV (0.3 eV) in the vicinity of the W-point
in Fig. 1(a) [Fig. 2(a)].

\begin{figure}[h]
\includegraphics[width=8cm]{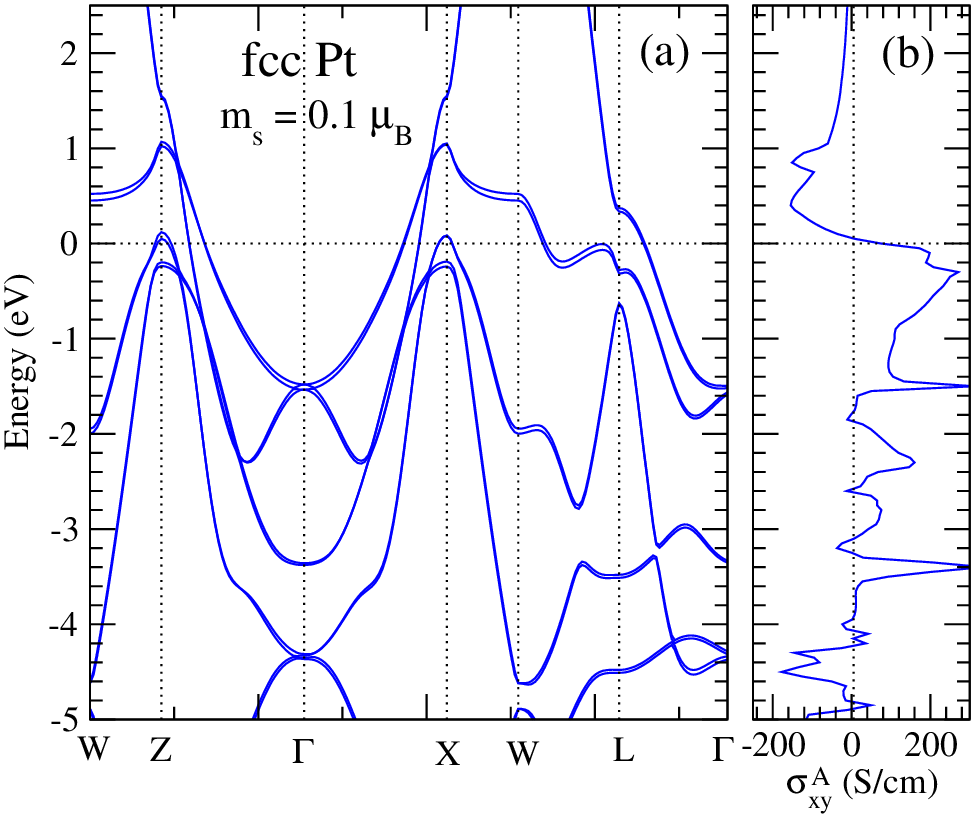}
\caption{\label{BS1} (Color online) (a) Relativistic band structure and (b) anomalous Hall
conductivity (AHC) of the magnetized platinum with a spin magnetic moment of 0.1 $\mu_B$/atom.
The horizontal dotted line at the zero energy indicates the Fermi level. 
}
\end{figure}

\begin{figure}[h]
\includegraphics[width=8cm]{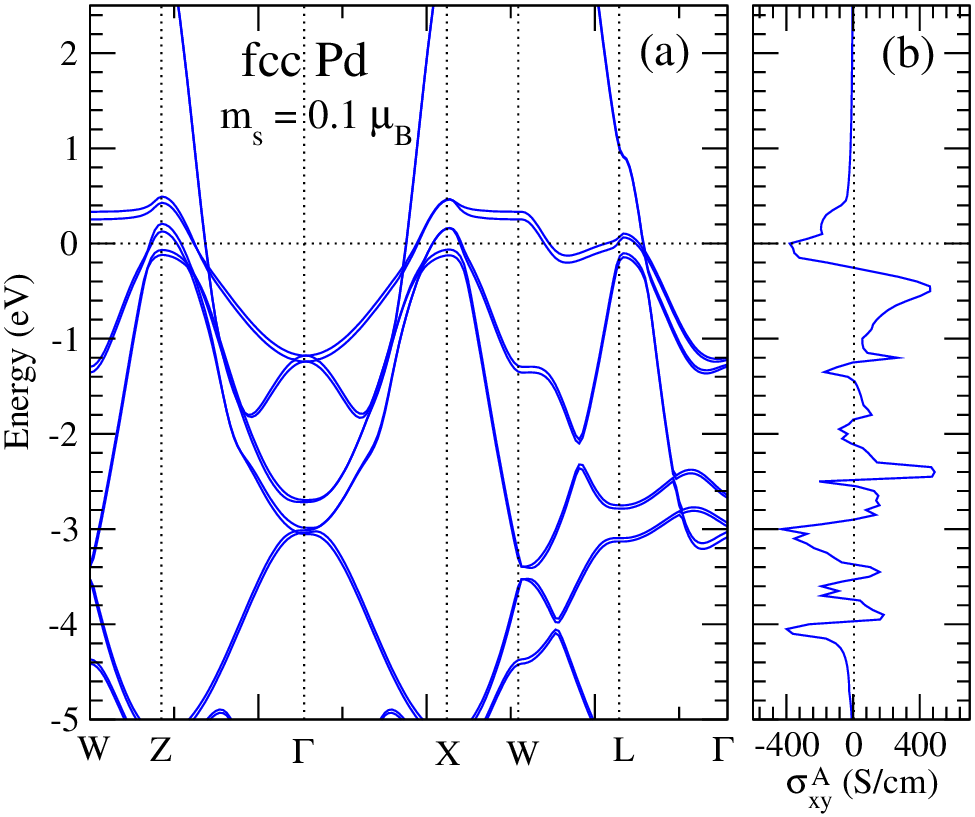}
\caption{\label{BS2} (Color online) (a) Relativistic band structure and (b) anomalous Hall
conductivity of the magnetized palladium with a spin magnetic moment of 0.1 $\mu_B$/atom.
The horizontal dotted line at the zero energy indicates the Fermi level.}
\end{figure}

\subsection{Anomalous Hall effect}

Figures 3 and 4 show the calculated AHC and ANC ($\alpha_{xy}^A$) as well as the exchange splitting ($\Delta_{ex}$) 
as a function of the induced spin magnetic moment ($m_s$) in platinum and palladium, 
respectivelty. $\Delta_{ex}$ refers to the splitting of the spin-up and spin-down
bands, and we calculate $\Delta_{ex}$ as the spin splitting of the scalar-relativistic 
bands above the Fermi level at the W-point [Figs. 1(a) and 2(a)]. First of all, it is clear 
from Figs. 3 and 4 that the calculated $\sigma_{xy}^A$ and $\Delta_{ex}$ increase monotonically with $m_s$. 
In fact, $\Delta_{ex}$ is almost perfectly proportional to $m_s$, while the amplitude of the $\sigma_{xy}^A$ 
increases linearly with $m_s$ for small $m_s$ values up to 0.30 and 0.25 $\mu_B$/atom
for Pt and Pd, respectively. 

Secondly, the AHC is large. In particular, the magnitude of the 
AHC per $\mu_B$ ($\sigma_{xy}^A/m_s$) for $m_s \le 0.25$ $\mu_B$/atom in Pt and Pd is, respectively, 
$\sim$790 and 3500 S/(cm$\cdot\mu_B$), being much larger than that of $\sim$360 S/(cm$\cdot\mu_B$) 
in iron\cite{Yao04}. Interestingly, the ratio $\sigma_{xy}^A/m_s$ for Pt is smaller than that
for Pd, indicating that the AHC in a proximity-induced ferromagnetic metal is not necessarily 
correlated with the SOC strength. Thirdly, the sign of the AHC in Pt is opposite to
that in Pd, being in good agreement with the recent experiments on the Pt/YIG and Pd/YIG
bilayers\cite{Zhou13}. 

\begin{figure}[h]
\includegraphics[width=7cm]{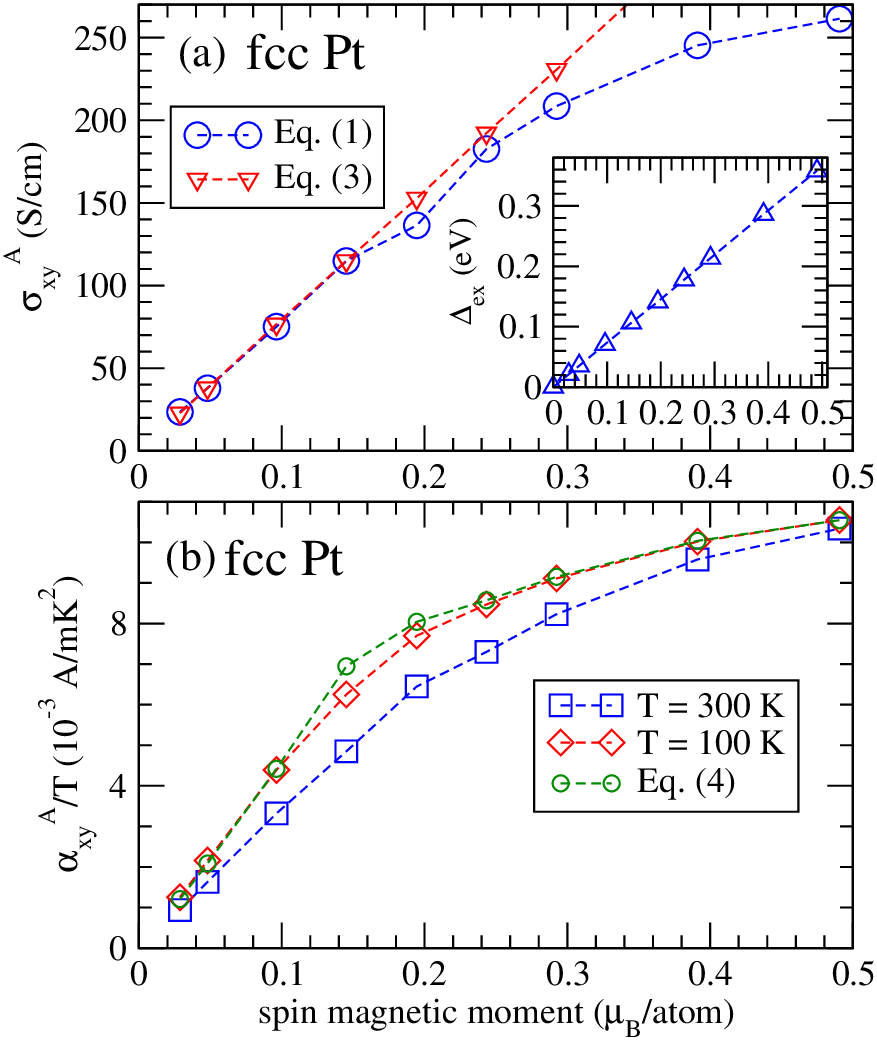}
\caption{\label{ANC1} (Color online) (a) Anomalous Hall conductivity ($\sigma_{xy}^A$)
and (b) anomalous Nernst conductivity ($\alpha_{xy}^A$) as a function of the induced spin magnetic
moment ($m_s$) in platinum. Exchange splitting ($\Delta_{ex}$) is displayed as a function
of $m_s$ in the inset in (a). In (b), $T$ denotes temperature.}
\end{figure}

\begin{figure}[h]
\includegraphics[width=7cm]{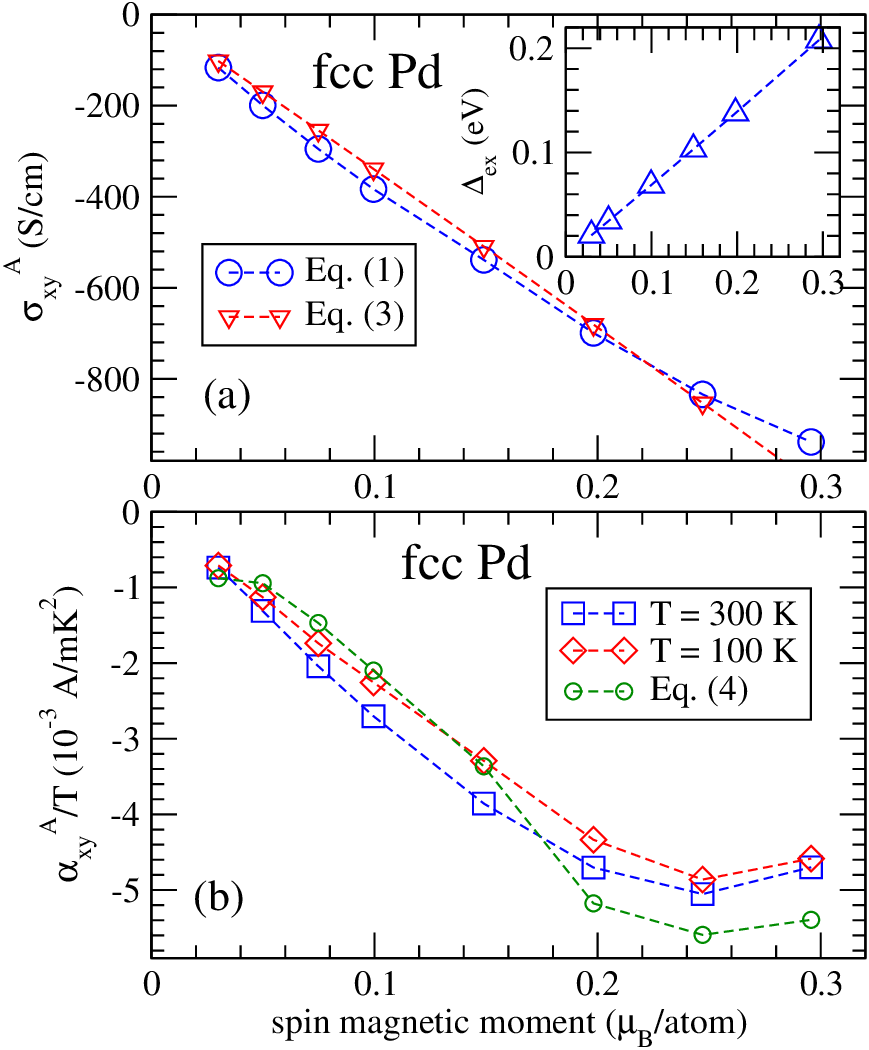}
\caption{\label{ANC2} (color online) (a) Anomalous Hall conductivity ($\sigma_{xy}^A$)
and (b) anomalous Nernst conductivity ($\alpha_{xy}^A$) as a function of the induced spin magnetic
moment ($m_s$) in palladium. Exchange splitting ($\Delta_{ex}$) is displayed as a function
of $m_s$ in the inset in (a). In (b), $T$ denotes temperature.}
\end{figure}

\subsection{Correlation between anomalous and spin Hall conductivities}
In order to gain insight into the key factors that determine the AHC in a magnetized nonmagnetic metal,
let us consider the two-current model to connect the conductivities for the different sorts of Hall effects.
Within the two-current model approximation, $\sigma_{xy}^{A}$ and $\sigma_{xy}^{S}$ can
be written as\cite{Tung12,Tung13,note1} $\sigma_{xy}^{A}(E) = \sigma_{xy}^{\uparrow}(E) + \sigma_{xy}^{\downarrow}(E)$ and
$-2\frac{\hbar}{e}\sigma_{xy}^{S}(E) = \sigma_{xy}^{\uparrow}(E) - \sigma_{xy}^{\downarrow}(E)$, where
$\sigma_{xy}^{\uparrow}$ and $\sigma_{xy}^{\downarrow}$ are the spin-up and spin-down Hall conductivities,
respectively. In a non-magnetic metal, the spin magnetic moment $m_s=0$ and thus, $\sigma_{xy}^{A}=0$. In the magnetized metal,
$\sigma_{xy}^{A}(E) = \sigma_{xy}^{\uparrow}(E-\frac{1}{2}\Delta_{ex}) + \sigma_{xy}^{\downarrow}(E+\frac{1}{2}\Delta_{ex})
\approx \sigma_{xy}^{\uparrow}(E) - \frac{1}{2}\Delta_{ex}\sigma_{xy}^{\uparrow}(E)'+\sigma_{xy}^{\downarrow}(E)
+\frac{1}{2}\Delta_{ex}\sigma_{xy}^{\downarrow}(E)' $, where $\Delta_{ex}$ is the exchange splitting and is proportional
to $m_s$, as shown in the inset in Fig. 3(a) and Fig. 4(a). 
Therefore, we find 
\begin{equation}
\sigma_{xy}^A(E_F) \approx \Delta_{ex}\frac{e}{\hbar}\sigma_{xy}^S(E_F)'.
\end{equation}
Equation (3) tells us that the AHC is proportional to the energy derivative of 
the spin Hall conductivity [$\sigma_{xy}^S$]
as well as the exchange splitting ($\Delta_{ex}$). Interestingly, the SOC strength does not appear 
explicitly in Eq. (3), in contrary to conventional wisdom. We notice that platinum and palladium
have similar AHC-versus-energy [$\sigma_{xy}^A(E)$] curves which have a prominent peak near
the $E_F$ (see Fig. 1 in both \cite{Guo08} and \cite{Guo09}).  
However, the $E_F$ falls on the up-hill side of the peak in Pt\cite{Guo08} but on the down-hill
side of the peak in Pd\cite{Guo09}, resulting in the positive $\sigma_{xy}^S(E_F)'$ for Pt and
negative $\sigma_{xy}^S(E_F)'$ for Pd. This, together with Eq. (3), naturally explains why both the calculated
and observed AHCs in Pt and Pd have opposite signs.

\begin{figure}[h]
\includegraphics[width=7cm]{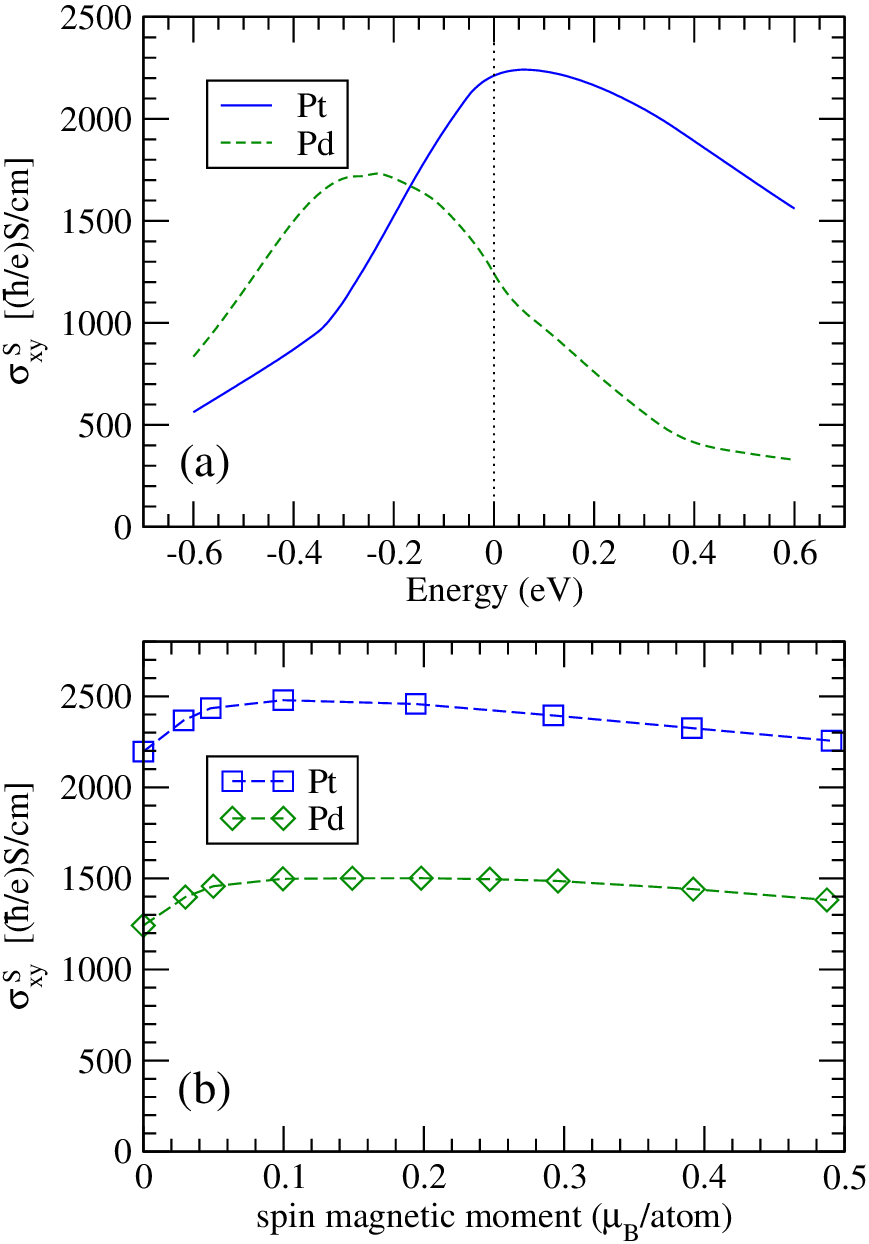}
\caption{\label{SHC} (Color online) (a) Spin Hall conductivity ($\sigma_{xy}^S$)
as a function of energy in nonmagnetic Pt and Pd metals. The vertical dotted line
at 0 eV indicates the Fermi energy ($E_F$). $\sigma_{xy}^S(E_F) = 2200$ and 1242 ($\hbar$/e)S/cm 
for Pt and Pd, respectively. $\sigma_{xy}^S(E_F)' = 1081$ and -4245 ($\hbar$/e)S/cm-eV 
for Pt and Pd, respectively. (b) Spin Hall conductivity as a function of the induced spin magnetic
moment ($m_s$) in magnetized Pt and Pd metals.}
\end{figure}
 
To examine quantitatively the validity of Eq. (3), here we repeat the calculations 
of the SHC for Pt\cite{Guo08} and Pd\cite{Guo09} but using the more accurate FLAPW method 
with the same computational details as described already in Sec. II. 
The calculated SHC for Pt and Pd as a function of energy is displayed in Fig. 5(a).
The $\sigma_{xy}^S$ at the $E_F$ is 2200 ($\hbar$/e)S/cm for Pt and 1242 ($\hbar$/e)S/cm 
for Pd, being in good agreement with the corresponding results 
calculated previously using the linear muffin-tin orbital method with the atomic sphere approximation.\cite{Guo08,Guo09}
We then evaluate numerically the energy derivative of the SHC using 
the $\sigma_{xy}^S(E)$ displayed in Fig. 5. We obtain $\sigma_{xy}^S(E_F)' = 1081$ 
and -4245 ($\hbar$/e)S/cm-eV for Pt and Pd, respectively.  Figures 3(a) and 4(a) also 
show the $\sigma_{xy}^A$ evaluated using Eq. (3) together with the calculated 
$\sigma_{xy}^S(E_F)'$ and $\Delta_{ex}$. It is clear that Eq. (3) holds very well 
for small $m_s$ up to $\sim$0.25 $\mu_B$/atom for Pt and Pd [Figs. 3(a) and 4(a)].

We have also calculated the SHC in the magnetized Pt and Pd metals. The calculated SHC
for Pt and Pd is shown as a function of the spin magnetic moment in Fig. 5(b). 
In both Pt and Pd, the SHC initially increases with $m_s$ up to $\sim$0.1 $\mu_B$/atom and then
decreases slowly as $m_s$ further increases [Fig. 5(b)]. Nevertheless, the SHC for both Pt and Pd
remains in the same order of magnitude all the way up to $m_s$ = 0.5 $\mu_B$/atom.

The validity of Eq. (3) may be understood at the microscopic level.
The two-current model can be derived from an approximation in which the spin-flipping part of the SOC
is ignored. The spin-conserving part of the SOC can still lead to nontrivial results on
the transverse transport coefficients.  
This non-flip approximation can be justified for crystals with inversion 
symmetry and in the limit of zero magnetization.  This is because that Kramer's theorem implies a two-fold 
degeneracy of the band structure at general $k$-points even in the presence of the SOC.  
The SOC term in the Hamiltonian, being symmetric under spatial inversion and time reversal, 
must behave as a constant within the degenerate space.  Therefore, it must also commute with the 
representation of the spin operator within the two-fold degenerate space.  
In the presence of a small magnetization, the degenerate bands are split to first order in the Zeeman energy 
according to the representation of the spin operator within each of the original degenerate space.  
Not being able to mix these split levels directly, the spin-flip part of the SOC term in the Hamiltonian can be 
safely discarded, because its residual effect must be of second order (in a process going out and back to the degenerate space).


\subsection{Anomalous Nernst effect}
Figures 3(b) and 4(b) indicate firstly that the anomalous Nernst conductivity $\alpha_{xy}^A$ 
increases monotonically with the spin moment $m_s$ for $m_s$ up to at least 
0.5 $\mu_B$/atom in Pt and for $m_s$ up to 0.25 $\mu_B$/atom in Pd. 
Like $\sigma_{xy}^A$, $\alpha_{xy}^A$ is approximately proportional to $m_s$
for $m_s \le \sim 0.20$ $\mu_B$/atom in both Pt and Pd. 
Secondly, the calculated $\alpha_{xy}^A$ is large, especially in Pt [Fig. 3(b)]. 
In fact, $\alpha_{xy}^A$ for Pt at $m_s \ge 0.15$ $\mu_B$/atom could be ten times 
larger than the intrinsic $\alpha_{xy}^A$ [$\alpha_{xy}^A$/T = $0.51\times10^{-3}$ A/(m-K$^2$) 
at $T = 293$ K] of iron\cite{Weis13}.
The magnitude of $\alpha_{xy}^A$ for Pd for $m_s \ge 0.05$ $\mu_B$/atom
is also several times larger than that of iron\cite{Weis13}. 

At low temperatures, Eq. (2) can be simplified as the Mott relation,
\begin{equation}
\alpha_{xy}^A = -\frac{\pi^2}{3}\frac{k_B^2T}{e}\sigma_{xy}^A(\mu)',
\end{equation}
which relates the ANC to the AHC. 
Therefore, it is not surprising that the magnetized platinum has a very large 
$\alpha_{xy}^A$ since the $E_F$ is located on the steep slope of
$\sigma_{xy}^A(E)$ [Fig. 1(b)], resulting in a large energy derivative 
of $\sigma_{xy}^A(E)$ at $E_F$. In Figs. 1(b) and 2(b), the $\alpha_{xy}^A$ calculated 
using the Mott relation [Eq. (4)] is displayed as a function of the induced $m_s$. 
Clearly, $\alpha_{xy}^A$ calculated directly [Eq. (2)] and using the Mott relation 
at $T = 100$  K are in good agreement with each other for Pt [see Fig. 3(b)]
and also for  $m_s \le 0.15$ $\mu_B$/atom for Pd [see Fig. 4(b)].
On the other hand, $\alpha_{xy}^A$ at $T = 300$ K calculated directly differs 
noticeably from that from the Mott relation, indicating that $T = 300$ K 
cannot be considered as a low temperature in this context.

Differentiating Eq. (3) and substituting the result into Eq. (4), we find 
\begin{equation}
\alpha_{xy}^A/T = -\frac{\pi^2}{3}\frac{k_B^2\Delta_{ex}}{\hbar}\sigma_{xy}^s(\mu)''.
\end{equation}
Eqs. (3) and (4) indicate that for small $m_s$, both the AHC and ANC are proportional to
the exchange-splitting. As mentioned above, the exchange-splitting is almost a perfect
linear function of $m_s$, and hence this explains why both the $\sigma_{xy}^A$ and $\alpha_{xy}^A$ 
are approximately proportional to $m_s$. Furthermore, this suggests that 
the $\sigma_{xy}^A$ and $\alpha_{xy}^A$ are proportional to each other for small $m_s$,
as shown in Fig. 3 and Fig. 4.
Therefore, we can rewrite Eq. (3) as 
\begin{equation}
\sigma_{xy}^A(E_F) \approx [\frac{e}{\hbar}\frac{\Delta_{ex}(m_s^0)}{m_s^0}\sigma_{xy}^S(E_F)']m_s = \beta m_s,
\end{equation}
where constant $\beta$ can be determined solely by first-principle calculations
for a certain spin moment $m^0_s$. In the present work, we find that $\beta= 788$ S/cm/$\mu_B$
for Pt and $\beta = -2921$ S/cm/$\mu_B$ for Pd.
Similarly, we can rewrite Eq. (4) as 
\begin{equation}
\alpha_{xy}^A/T \approx -[\frac{\pi^2}{3}\frac{k_B^2\Delta_{ex}(m_s^0)}{\hbar m_s^0}\sigma_{xy}^S(E_F)'']m_s = \gamma m_s.
\end{equation}
And using the results of the first-principle calculations, we obtain that
constant $\gamma = 0.034$ A/(m-K$^2\mu_B$) for Pt and $\gamma = -0.027$ A/(m-K$^2\mu_B$) for Pd.

\section{Closing remarks}
Recently, the possible magnetic proximity-induced spin moment in Pt films in the Pt/YIG bilayers was
measured by magnetic x-ray circular dichroism experiments\cite{Lu13}, and $m_s$ was
found to be 0.054 $\mu_B$/atom at 300 K and 0.076 $\mu_B$/atom at 20 K. Using $m_s = 0.05$ $\mu_B$/atom
together with Eqs. (6) and (7), we can estimate the intrinsic AHC and ANC for the Pt film to be
$\sigma_{xy}^A$  =  40 S/cm and $\alpha_{xy}^A$ =  0.51 A/(m-K$^2$) ($T = 300$ K).  
The anomalous Seebeck coefficient $E_y/(-\partial_xT) = \rho_{xx}(\alpha_{xy}-S\sigma_{xy})$
where $S = \alpha_{xx}/\sigma_{xx}$ is the ordinary Seebeck coefficient. 
At $T = 300$ K, $\rho_{xx} = 10.8$ $\mu\Omega$cm and $S = -11.28$ $\mu$V/K (see Refs. \onlinecite{Landa} and
\onlinecite{Landb}). Resultantly, $E_y/(-\partial_xT) = 0.058 \mu$V/K. 
Using the sample sizes and the temperature gradient in the Pt/YIG bilayers\cite{Qu13,Kikk13},
one would obtain the Hall voltage due to the ANE in the order of $\sim$0.1 $\mu$V, being comparable 
with the Hall voltage ($\sim$0.1 $\mu$V in Au/YIG and $\sim$1.0 $\mu$V in Pt/YIG)
produced by the spin Seebeck effect via the inverse spin Hall effect. 

\section*{Acknowledgments}
G.Y.G. acknowledges support from the Ministry of Science and Technology, the Academia Sinica Thematic 
Research Program and NCTS of Taiwan, and thanks Shang-Fan Lee and Shiming Zhou for stimulating discussions. G.Y.G.
also acknowledges partial support from NBRPC (No. 2012CB921300 and No. 2013CB921900), and NSFC (No. 91121004)
during his visit at Peking University. Q.N. was supported in part by DOE-DMSE (No. DE-FG03-02ER45958)
and the Welch Foundation (No. F-1255).



\begin{references}
\bibitem{note1} Note that in Refs. \onlinecite{Tung12} and \onlinecite{Tung13}, 
$2\frac{\hbar}{e}\sigma_{xy}^{S} = \sigma_{xy}^{\uparrow} - \sigma_{xy}^{\downarrow}$ 
in Eq. (5) and Eq. (2b), respectively, should be 
$-2\frac{\hbar}{e}\sigma_{xy}^{S} = \sigma_{xy}^{\uparrow} - \sigma_{xy}^{\downarrow}$.
Consequently, the signs of the spin-up and spin-down Hall conductivities 
($\sigma_{xy}^{\uparrow}$ and $\sigma_{xy}^{\downarrow}$) as well as the spin polarization of the Hall
current ($P^H$) listed in Tables I and II in Refs. \onlinecite{Tung12} and \onlinecite{Tung13} should
be reversed.
\end{references}
\end{document}